# Online Rotor Resistance Adaptation Of Induction Motor Drive

Yuva sudhakar          P.M.Tiwari

**Depart. Electrical and Electronic Engineering,**

**Amity University, India.**

*Abstract:* In this paper an adaptive control systems scheme is used to update the rotor resistance of an induction motor drive. Rotor resistance of the induction motor drive is dependent on the temperature where it is installed for one Induced Draft Fan (I.D Fan) in one of the 500 MW generating thermal power plant. The rotor resistance adaptation is of induction motor drive is achieved through Indirect Field Oriented. The desired value of the rotor flux along the q - axis should ideally be zero. This condition acts as a reference model for the proposed Adaptive Control scheme. Inductance does not change with the temperature. Therefore, the only parameter which changes with the temperature in the adjustable model is the rotor resistance. This resistance is adjusted such that the flux along the q - axis is driven to zero. The proposed method effectively adjusts the rotor resistance on-line and keeps the machine field oriented. The effectiveness of the proposed scheme is tested through simulation and is experimentally verified.

## 1. Introduction.

An off-line tuning scheme presented in [l] is used to find the initial values of the rotor inductance and resistance at ambient temperature. The rotor inductance is a function of the current and speed while the rotor resistance and hence the rotor time constant $(T = L_r/R_r)$ varies substantially when the temperature changes. This results in the detuning of the machine. Therefore, an on-line adaptation of the rotor resistance and hence $T_r$ is necessary to keep the machine field oriented. Many on-line rotor resistance identification schemes have been designed [4]-[6]. Some fuzzy logic based techniques [7]-[9] have been proposed to overcome the detuning. Some fuzzy logic based techniques [7]-[9] have been proposed. To overcome the detuning. The current model flux observer in the synchronous frame is found to be most appropriate for on-line rotor resistance adaptation The rotor flux equation in the synchronous frame acts as an adjustable model. This model is modified through the rotor resistance adaptation, which will then force the flux along the q - axis to zero.

## 2. Modelling of induction motor.

The induction machine d-q or dynamic equivalent circuit is shown in Fig. 1. One of the most popular induction motor models derived from this equivalent circuit is Krause's model. According to his model, the modelling equations in flux linkage form are as follows:

$$\frac{dF_{qs}}{dt} = \omega_b \left[ v_{qs} - \frac{\omega_e}{\omega_b} F_{ds} + \frac{R_s}{x_{ls}} \left( F_{mq} + F_{qs} \right) \right] \quad (1)$$

$$\frac{dF_{ds}}{dt} = \omega_b \left[ v_{ds} + \frac{\omega_e}{\omega_b} F_{qs} + \frac{R_s}{x_{ls}} \left( F_{md} + F_{ds} \right) \right] \quad (2)$$

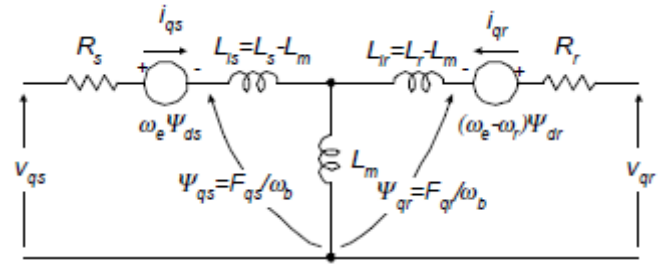

(a)

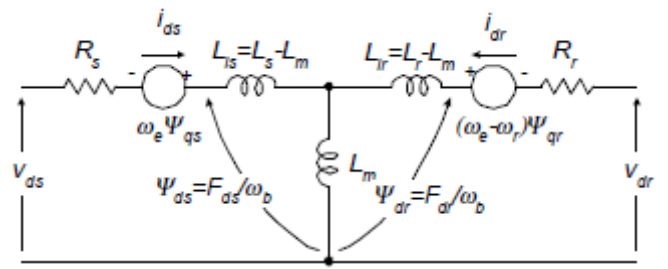

(b)

Fig. 1. Dynamic or d-q equivalent circuit of an induction machine.

$$\frac{dF_{qr}}{dt} = \omega_b \left[ v_{qr} - \frac{(\omega_e - \omega_r)}{\omega_b} F_{dr} + \frac{R_r}{x_{lr}} (F_{mq} - F_{qr}) \right] \quad (3)$$

$$\frac{dF_{dr}}{dt} = \omega_b \left[ v_{dr} + \frac{(\omega_e - \omega_r)}{\omega_b} F_{qr} + \frac{R_r}{x_{lr}} (F_{md} - F_{dr}) \right] \quad (4)$$

$$F_{mq} = x_{ml}^* \left[ \frac{F_{qs}}{x_{ls}} + \frac{F_{qr}}{x_{lr}} \right] \quad (5)$$

$$F_{md} = x_{ml}^* \left[ \frac{F_{ds}}{x_{ls}} + \frac{F_{dr}}{x_{lr}} \right] \quad (6)$$

$$i_{qs} = \frac{1}{x_{ls}} (F_{qs} - F_{mq}) \quad (7)$$

$$i_{ds} = \frac{1}{x_{ls}} (F_{ds} - F_{md}) \quad (8)$$

$$i_{qr} = \frac{1}{x_{lr}} (F_{qr} - F_{mq}) \quad (9)$$

$$i_{dr} = \frac{1}{x_{lr}} (F_{dr} - F_{md}) \quad (10)$$

$$T_e = \frac{3}{2} \left( \frac{p}{2} \right) \frac{1}{\omega_b} (F_{ds} i_{qs} - F_{qs} i_{ds}) \quad (11)$$

$$T_e - T_L = J \left( \frac{2}{p} \right) \frac{d\omega_r}{dt} \quad (12)$$

where $d$ : direct axis,
$q$ : quadrature axis,
$s$ : stator variable,
$r$ : rotor variable,
$Fij$ is the flux linkage ($i=q$ or $d$ and $j=s$ or $r$),
$vqs, vds$ : $q$ and $d$–axis stator voltages,
$vqr, vdr$ : $q$ and $d$–axis rotor voltages,
$Fmq, Fmd$ : $q$ and $d$ axis magnetizing flux linkages,
$Rr = 0.$ : rotor resistance,
$Vd = 260$volts.
$Rs = 0.6$ ohms : stator resistance,
$Xls=1.9mH$: stator leakage reactance ($\omega eLls$),
$Xlr=1.9mH$ : rotor leakage reactance ($\omega eLlr$),
$iqs, ids$ : $q$ and $d$–axis stator currents,
$iqr, idr$ : $q$ and $d$–axis rotor currents,
$p = 6$ : number of poles,
$J=3$ : moment of inertia,
$Te$ : electrical output torque,
$TL$(or $Tl$) : load torque,
$\omega e$ : stator angular electrical frequency,
$\omega b$ : motor angular electrical base frequency,
$\omega r$ : rotor angular electrical speed.

a. **o-n conversion block**

The transformation is represented as follows

$$\begin{bmatrix} v_{an} \\ v_{bn} \\ v_{cn} \end{bmatrix} = \begin{bmatrix} +\frac{2}{3} & -\frac{1}{3} & -\frac{1}{3} \\ -\frac{1}{3} & +\frac{2}{3} & -\frac{1}{3} \\ -\frac{1}{3} & -\frac{1}{3} & +\frac{2}{3} \end{bmatrix} \begin{bmatrix} v_{ao} \\ v_{bo} \\ v_{co} \end{bmatrix} \quad (13)$$

b. **abc-syn and syn-abc conversion**

To convert 3-ph voltages of stationery frame to 2- ph synchronously rotating frame and vice versa.

$$\begin{bmatrix} v_{qs}^s \\ v_{ds}^s \end{bmatrix} = \begin{bmatrix} 1 & 0 & 0 \\ 0 & -\frac{1}{\sqrt{3}} & \frac{1}{\sqrt{3}} \end{bmatrix} \begin{bmatrix} v_{an} \\ v_{bn} \\ v_{cn} \end{bmatrix} \quad (13)$$

$$\begin{cases} v_{qs} = v_{qs}^s \cos\theta_e - v_{ds}^s \sin\theta_e \\ v_{ds} = v_{qs}^s \sin\theta_e + v_{ds}^s \cos\theta_e \end{cases} \quad (14)$$

$$\begin{bmatrix} i_a \\ i_b \\ i_c \end{bmatrix} = \begin{bmatrix} 1 & 0 \\ -\frac{1}{2} & -\frac{\sqrt{3}}{2} \\ -\frac{1}{2} & \frac{\sqrt{3}}{2} \end{bmatrix} \begin{bmatrix} i_{qs}^s \\ i_{ds}^s \end{bmatrix} \quad (15)$$

$$\begin{cases} i_{qs}^s = v_{qs}\cos\theta_e + v_{ds}\sin\theta_e \\ i_{ds}^s = -v_{qs}\sin\theta_e + v_{ds}\cos\theta_e \end{cases} \quad (16)$$

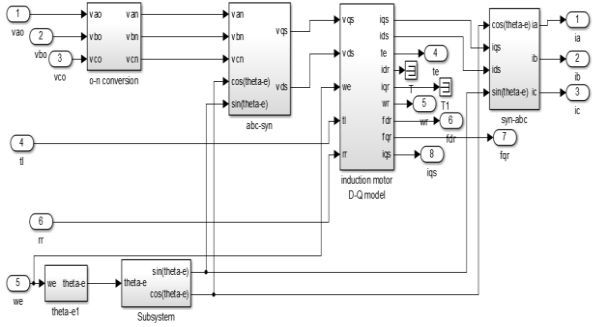

Fig-2 Simulink model of induction motor.

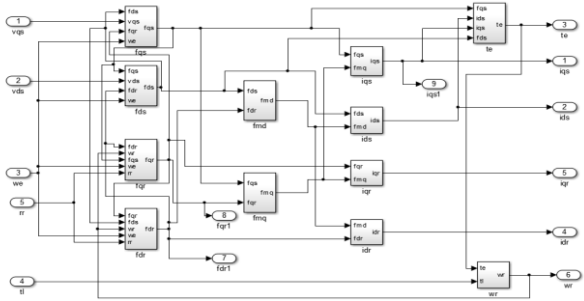

Fig-3 D-Q model of induction motor drive

### 3. Current observer design

An observer or estimator has an integration process model that estimates the desired state. An observer or estimator has an integration process model that estimates the desired state. the open loop observer is essentially a real-time simulation of the physical process. Current Model Open Loop Flux Observer are analyzed to explore which one is the more suitable for rotor resistance adaptation. close loop observer is constructed with the reference model which is $\lambda_{qr} = 0$

The current model open loop flux observer can estimate the rotor flux directly from a knowledge of the stator current, slip speed ($\omega_{sl}$)rotor inductance *(Lr )* and rotor resistance *(Rr)*. This model is represented in Fig-4.

$$p\hat{\lambda}_{dr} = \frac{L_m}{T_r}i_{ds} - \frac{1}{T_r}\hat{\lambda}_{dr} + \omega_{sl}\hat{\lambda}_{qr} \quad (17)$$

$$p\hat{\lambda}_{qr} = \frac{L_m}{T_r}i_{qs} - \frac{1}{T_r}\hat{\lambda}_{qr} - \omega_{sl}\hat{\lambda}_{dr} \quad (18)$$

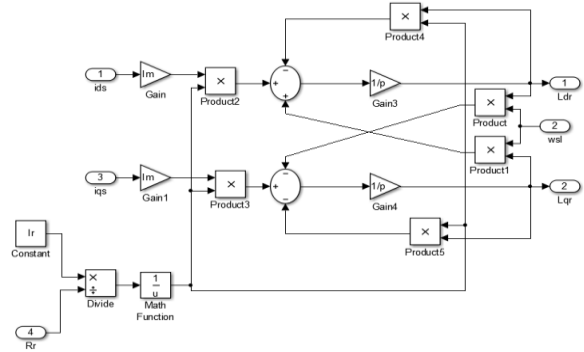

Fig-4 current model flux observer for rotor.

### 4. Online rotor resistance adaptation scheme

The flux observers reviewed above are mostly used, in the stationary frame of reference, either for direct field orientation control or for sensor less control. The observer structure is designed, in the synchronous frame of reference instead of stationary frame of reference, for an IFO drive system with a speed sensor on the motor shaft.

i. The flux is aligned along the *d - axis* of the machine and the flux along the *q - axis* of the machine is zero that is $\lambda_{qr} = 0$.

ii. Rotor resistance in the induction motor changes with the temperature in the current model flux observer.

iii. The rotor speed is obtained from shaft encoder.

The only unknown parameter in the system is rotor resistance $R_r$. If $R_r$ is adjusted the flux calculated from the adjustable model will be equal to the reference model. When the temperature changes $R_r$ deviates from the original value, which then drives A,, away from zero. This error in the flux can be forced back to zero by adjusting $R_r$.

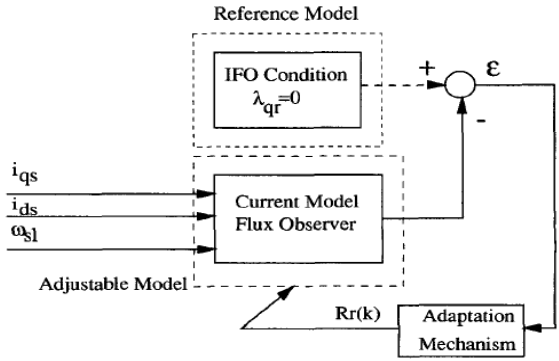

Fig-5. Rotor resistance adaptation through MRAC Scheme.

The flux error equation can be written as

$$\epsilon = \lambda_{qr(ref)} - \hat{\lambda}_{qr} \quad (19)$$

A PI controller is used for the rotor resistance adaptation.

$$R_r(k) = R_r(k-1) + \Delta R_r(k) \quad (20)$$

Where

$$\Delta R_r(k) = K_p \epsilon(k) + K_i \epsilon(k) T \quad (21)$$

## 5. Simulation results

The rotor resistance is roughly known at this temperature which can be used as the initial value to start the controller. When the machine is running its temperature changes, which then changes the rotor resistance. The rotor resistance needs to be adapted to keep the machine field oriented. To check the effectiveness of the proposed algorithm, the rotor resistance is set to a value lower than its ambient one (0.412). In this experiment, a step command of 250 rpm is applied and a PI controller is used for the speed control. The gains of PI controller are kept same to validate the effectiveness of the proposed scheme. The machine is operated under three different conditions.

- The command rotor resistance is set to one half of its original value.
- The command rotor resistance is set to one fourth of its original value.
- The adaptation is applied with the rotor resistance initialized at one fourth of its original value.

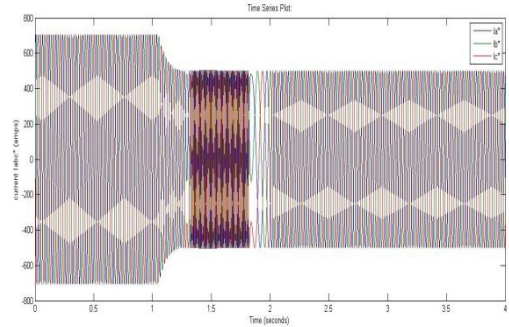

Fig-6. Currents (Iabc*) amperes

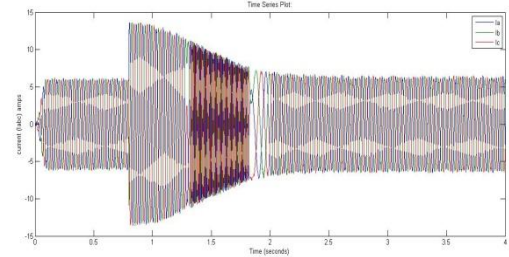

Fig-7 currents (iabc)

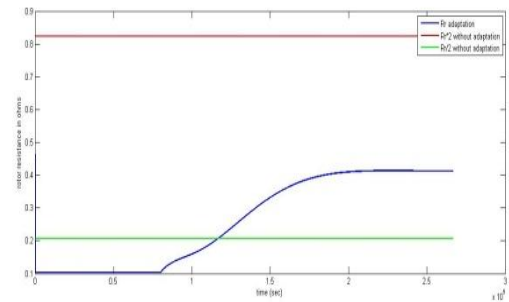

Fig-8. Resistance adaptation.

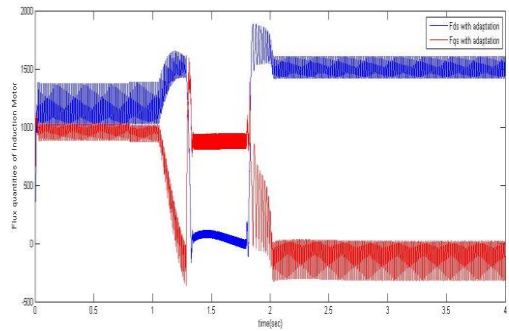

Fig-9. Stator Flux along d and q- axis.

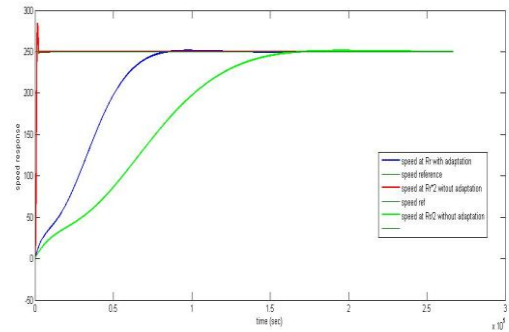

Fig-10. Speed response of induction motor.

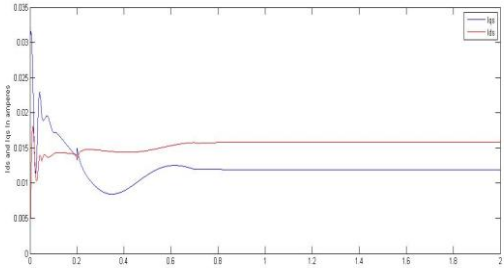
Fig-11. Ids and Iqs in amperes.

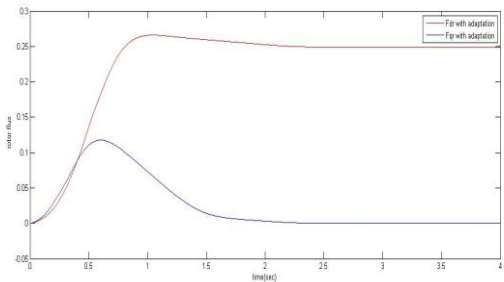
Fig-12. Rotor flux along d and q axis.

## 6. Conclusion

The results shows that the scheme provides the adaptation of Rr = 0.412 in 0.6 sec. This scheme is based on a current model flux observe rotating at synchronous speed. The proposed scheme works effectively to tune the rotor resistance to its correct value and keeps the machine field oriented.

## 7. Reference

[1] Habib-ur Rehman, Adnan Derdiyok, Mustafa K. Guven, Longya Xu "An MRAS Scheme for On-line Rotor Resistance Adaptation of an Induction Machine"

[2] P. L. Jansen and R. D. Lorenz, D. W. Novotny, " Observer-based direct field orientation: analysis and comparison of alternative methods", IEEE Transaction on Industry Applications, Vol. 30, No 4, pp. 945-953 July./Aug. 1994.

[3] L. Zhen and L. Xu, "On-line fuzzy tuning of indirect field oriented induction machine drives", in the Proceedings of IEEE APEC'96 Conference, San Jose, California, March 1996.

[4] ] H. Sugimoto, S. Tamai, "Secondary resistance identification of an induction motor applied model reference adaptive system and its characteristics" , IEEE Transaction on Industry Applications, Vol. IA-23, pp. 296-303, 1987.

[5] T. Matsuo, T. A. Lipo, "A rotor parameter identification scheme for vector-controlled induction machine motor drives" , IEEE Ransaction on Industry Applications, Vol. IA-21, pp. 624-632, May/June 1985.